\documentclass{pasj00}
\begin{document}
\SetRunningHead{K. Sugiyama et al.}{Synchronized Variability
of the Cep~A methanol maser at 6.7~GHz}
\Received{2008/03/08}%%{yyyy/mm/dd}
\Accepted{2008/04/27}%%{yyyy/mm/dd}

\title{A Synchronized Variation of the 6.7~GHz
Methanol Maser in Cepheus A}

%%% begin:list of authors
%%% \author{Koichiro \textsc{Sugiyama}}
%%% \affil{Graduate school of Science and Engineering,
%%%Yamaguchi University, Yoshida, Yamaguchi 753-8512}

%%% \author{B-Firstname \textsc{B-Familyname}}
%%% \affil{B-Address of Institute}\email{bbbbb@xxx.xxx.xx.xx}
%%% \and
%%% \author{C-Firstname {\sc C-Familyname}}
%%% \affil{C-Address of Institute}\email{ccccc@xxx.xxx.xx.xx}
%%% end:list of authors

%%% Please use the following style in case that sorting by 
%%% affilation is impossible. 
%
\author{Koichiro \textsc{Sugiyama},\altaffilmark{1}
Kenta \textsc{Fujisawa},\altaffilmark{1,2} 
Akihiro \textsc{Doi},\altaffilmark{3}
Mareki \textsc{Honma},\altaffilmark{4,5}
Yasuko \textsc{Isono},\altaffilmark{1}\\
Hideyuki \textsc{Kobayashi},\altaffilmark{4,6}
Nanako \textsc{Mochizuki},\altaffilmark{3}
and
Yasuhiro \textsc{Murata}\altaffilmark{3,7}
}
  \altaffiltext{1}{Graduate school of Science and Engineering,
Yamaguchi University,\\1677-1 Yoshida, Yamaguchi, Yamaguchi 753-8512}
  \altaffiltext{2}{Department of Physics, Faculty of Science,
Yamaguchi University,\\1677-1 Yoshida, Yamaguchi, Yamaguchi 753-8512}
  \altaffiltext{3}{The Institute of Space and Astronautical Science,
Japan Aerospace Exploration Agency,\\3-1-1 Yoshinodai, Sagamihara,
Kanagawa 229-8510}
  \altaffiltext{4}{VERA Project, National Astronomical Observatory
of Japan, 2-21-1 Osawa, Mitaka, Tokyo 181-8588}
  \altaffiltext{5}{Department of Astronomical Science, Graduate
University for Advanced Studies,\\2-21-1 Osawa, Mitaka, Tokyo 181-8588}
  \altaffiltext{6}{Mizusawa VERA Observatory, 2-12 Hoshigaoka,
Mizusawa, Iwate 023-0861}
  \altaffiltext{7}{Department of Space and Astronautical Science,
The Graduate University for Advanced Studies,\\3-1-1 Yoshinodai,
Sagamihara, Kanagawa 229-8510}
\email{m005wa@yamaguchi-u.ac.jp}
%%% \email{ddddd@xxx.xxx.xx.xx}
%%% \email{eeeee@xxx.xxx.xx.xx}
%%% \altaffiltext{2}{Address of Institute}

%% `\KeyWords{}' always has to be placed before `\maketitle'.
%%% Do NOT move this preamble from here!

\maketitle

\begin{abstract}
We present the results of daily monitoring of 6.7~GHz methanol maser
in Cepheus~A (Cep~A) using Yamaguchi~32-m radio telescope as well
as the results of imaging observations conducted with the JVN
(Japanese VLBI Network).
We indentified five spectral features, which are grouped into
red-shifted ($-$1.9 and $-$2.7~km~s$^{-1}$) and blue-shifted
($-$3.8, $-$4.2, and $-$4.9~km~s$^{-1}$),
and we detected rapid variabilities of these maser 
features within a monitoring period of 81~days.
The red-shifted features decreased in flux density to 50{\%} of its
initial value, while the flux density of the blue-shifted features
rapidly increased within a 30~days.
The time variation of these maser features showed two remarkable
properties; synchronization and anti-correlation between
the red-shifted and the blue-shifted.
The spatial distribution of the maser spots obtained by the JVN
observation showed an arclike structure with a scale of $\sim$1400~AU,
and separations of the five maser features were found to be
larger than 100~AU.
The absolute position of the methanol maser was also obtained based
on the phase-referencing observations, and the arclike structure
were found to be associated with the Cep~A-HW2 object, with the
elongation of the arclike structure nearly perpendicularly to the
radio continuum jet from the Cep~A-HW2 object.
These properties of the masers, namely, the synchronization of flux
variation, and the spectral and spatial isolation of features,
suggest that the collisional excitation by shock wave from a common
exciting source is unlikely.
Instead, the synchronized time variation of the masers can be
explained if all the maser features are excited by infrared
radiation from dust which is heated by a common exciting
source with a rapid variability.
\end{abstract}

%%% ******************** %%%
%%% ***  Introduction ** %%%
%%% ******************** %%%

\section{Introduction}\label{section:introduction}
The $5_{1}\rightarrow6_{0}A^{+}$
methanol maser transition at 6.7~GHz
has the strongest flux densities among methanol maser lines.
The methanol maser emission is thought to be produced
by radiative excitation in an infrared radiation field,
which is formed by dust near the protostar
with a temperature of $\sim$100-200~K
(\cite{1997MNRAS.288L..39S}; \cite{2001ApJ...554..173S};
\cite{2005MNRAS.360..533C}).
Some sources, however, are associated with
shocked gas (\cite{1998MNRAS.301..640W}; \cite{2003MNRAS.341..277D};
\cite{2004MNRAS.351..779D}).
Collisional excitation by shock wave
may produce the methanol maser emission.

A number of 6.7~GHz methanol masers have been found
to exhibit long-term variability.
\citet{1995MNRAS.272...96C} found that the sources
vary on a time-scale of several months.
\citet{2000A&AS..143..269S} showed that about 65{\%} of
methanol maser in their sample exhibit moderate or strong
variability on time-scales of about four and eight years.
A long-term monitoring program to investigate the variability
of 54 methanol maser sources at 6.7~GHz has been conducted
using the Hartebeesthoek 26-m radio telescope
by \citet{2004MNRAS.355..553G}.
They divided the variable sources into six types according to
the behavior of the variability.
For G9.62$+$0.20E in their sample, the periodic variations
associated with massive star formation were discovered
for the first time \citep{2003MNRAS.339L..33G}.
The Very Long Baseline Interferometer (VLBI) monitoring
at seven epochs over three months
for G9.62$+$0.20E was conducted within the period of
the time variation \citep{2005MNRAS.356..839G}.
No appearance of new spots and no change in morphology was found,
suggesting that the flares were caused by a change
in either the seed or pump photon levels.
They proposed a binary system as an origin of the periodic flares.
\citet{1996MNRAS.280..868M} continuously observed G~351.78$-$0.54,
and found flares at least seven times with
time delays in the range 10-35~days between the variability
of red-shifted spectral feature and blue-shifted feature.
They discussed the delay, which can possibly be explained in terms of
light-crossing time due to the spatial distribution of spots.

We have conducted daily monitoring for some
maser sources to investigate a short time-scale variability
with Yamaguchi~32-m radio telescope.
Cepheus~A (Cep~A) was observed as one of our sample sources.
Cep~A is a CO condensation at a distance of 725~pc
\citep{1957ApJ...126..121J} and some radio continuum sources
were detected.
Methanol masers are associated with Cep~A-HW2 defined by
\citet{1984ApJ...276..204H}, which is the brightest
radio continuum source detected in the region.
The HW2 object has a radio continuum jet along
a position angle (PA) of $\sim$\timeform{45D}
(\cite{1994ApJ...430L..65R}; \cite{1995MNRAS.272..469H};
\cite{1996ApJ...457L.107T}, \yearcite{1998ApJ...509..262T};
\cite{2006ApJ...638..878C}).
Based on submillimeter observations of
both dust and CH$_3$CN line emissions,
a flattened disklike structure was found \citep{2005Natur.437..109P}.
The structure has a size of about 1000~AU
and is perpendicular to the jet.
Recently, it is revealed that NH$_3$ and SO$_2$ line emissions
coincided with the CH$_3$CN disk,
although the SO$_2$ structure was about 2 times smaller
(\cite{2007ApJ...666L..37T}; \cite{2007ApJ...661L.187J}).
The 22.2~GHz water masers observed by \citet{1996ApJ...457L.107T}
were distributed in an elongated structure perpendicular
to the radio jet, which possibly trace the circumstellar disk
around the HW2 object.
The water maser has a spatial distribution similar to the SO$_2$ disk.
The methanol maser at 6.7~GHz in Cep~A shows variability.
It was 1420~Jy in 1991 \citep{1991ApJ...380L..75M},
but 815~Jy in 1999 \citep{2000A&AS..143..269S}.
\citet{2003AJ....126.1967G} has found that the
amplitude ratio of the lines at a radial velocity of
$-$3.8~km~s$^{-1}$ and $-$4.2~km~s$^{-1}$
varies from 1.4 to 0.8 over two years.
The spatial distribution of the masers at 6.7~GHz for Cep~A
have been already reported \citep{2008PASJ..60..23P}
with the Japanese VLBI Network (JVN; \cite{2006astro.ph.12528D}).
However, the image quality was not enough to investigate
the relationship the spectral features and the spatial distribution.

In this paper, we present the spectral variability with the
spectral monitoring observations and a new VLBI map.
In section 2, we describe the details of these observations
and data reduction,
and the results are presented in section~3.
In section 4, we discuss interpretations of the rapid
variability and excitation mechanism of this maser in Cep~A.

%%% ******************** %%%
%%% ***  Observation  ** %%%
%%% ******************** %%%

\section{Observations and Data Reduction}\label{section:observation}
\subsection{Spectral Monitoring}\label{section:single}
The daily monitoring program with Yamaguchi 32-m radio telescope
was made from August 4 (corresponding to the days of year
(DOY) 216) to October 24 (DOY 297), 2007.
The full-width at half maximum (FWHM) of the beam
is 5~arcmin at 6.7~GHz.
The pointing error of the antenna is smaller than 1~arcmin.
The spectrometer consists of the IP-VLBI system
\citep{2003ASPC..306..205K} and a software spectrometer.
Both left and right circular polarizations
were recorded with 2-bit sampling.
The recorded data with a bandwidth of 4~MHz, covering a velocity range
of 180~km~s$^{-1}$, were divided into 4096 channels, yielding
a velocity resolution of 0.044~km~s$^{-1}$.
The integration time is 14 minutes until DOY 244,
and then 10 minutes from DOY 245.
The rms noise level was typically 1.2~Jy and 1.4~Jy with
an integration time of 14 min and 10 min, respectively.
An amplitude and a gain calibration was performed by
measuring the system noise temperatures by injecting the signal
from noise-sources with known temperatures.
The accuracy of the calibration was estimated to be 10{\%}.
The stability of the system was checked by daily monitoring
G12.91$-$0.26 methanol maser emission, which showed
relatively small variability in the sample of
\citet{2004MNRAS.355..553G}.

\subsection{VLBI Observation}\label{section:VLBI}
A VLBI observation at 6.7~GHz of Cep~A was made
on September 9 2006 from 15:00 to 22:00 UT with four telescopes
(Yamaguchi 32-m, Usuda 64-m, VERA-Mizusawa 20-m,
and VERA-Ishigaki 20-m) of the JVN.
The projected baselines covered from 9~M$\lambda$
(Usuda--Mizusawa) to 50~M$\lambda$ (Mizusawa--Ishigaki),
corresponding to the fringe spacing of 23~mas and 4.1~mas,
respectively.
Left-circular polarization was received at Yamaguchi and Usuda
stations, while linear polarization was received at Mizusawa
and Ishigaki stations.
A special amplitude calibration for different polarizations
was made by the same procedure as described in Sugiyama et al. (2008).
The data were recorded on magnetic tapes using the VSOP-terminal
system at a data rate of 128~Mbps with 2-bit quantization
and 2 channels, and correlated
at the Mitaka FX correlator \citep{Shibata_etal.1998}.
From the recorded 32~MHz bandwidth,
2~MHz (6668-6670~MHz) was divided into 512 channels for maser
reduction, yielding a velocity resolution of 0.176~km~s$^{-1}$.
This is four times broader than that of the single-dish observation.

A continuum source J2302$+$6405 (\timeform{2D.19} from Cep~A),
whose coordinate is known with an accuracy of 0.62~mas
in the third VLBA Calibrator Survey (VCS3) catalog
\citep{2005AJ....129.1163P}, was used as a phase reference calibrator.
We alternately observed the Cep~A methanol maser
and the continuum source with switching observational mode
with a cycle of 5~min (2~min on Cep~A and 1.6~min
on the continuum source).
Cep~A was sometimes continuously
observed for more than 30-min to improve the UV-coverage.
The total on-source times were 2.8 hours and 0.7 hours
for Cep~A and J2302$+$6405, respectively.
Bright continuum sources, 3C~454.3 and 3C~84, were also observed
every one and half hours for clock and bandpass calibration.
The synthesized beam has a size of 9.0~$\times$~3.5~mas
with a PA of $-$\timeform{70D}.

The data were reduced using the Astronomical Image Processing
System (AIPS: \cite{Greisen2003})
and the Difmap software \citep{Shepherd1997}.
The visibilities of all velocity channels
were phase-referenced to the reference maser spot
at LSR velocity of $-$2.64~km~s$^{-1}$, which is the brightest spot.
The absolute coordinate of the spot of
$-$2.64~km~s$^{-1}$ was obtained by applying the solution of phase
for the velocity channel to the phase reference source J2302$+$6405.

%%% ******************** %%%
%%% ***  Results      ** %%%
%%% ******************** %%%

\section{Results}\label{section:result}
\subsection{Spectral Variability}\label{section:vari}
\begin{figure}[htbp]
\begin{center}
\FigureFile(80mm,80mm){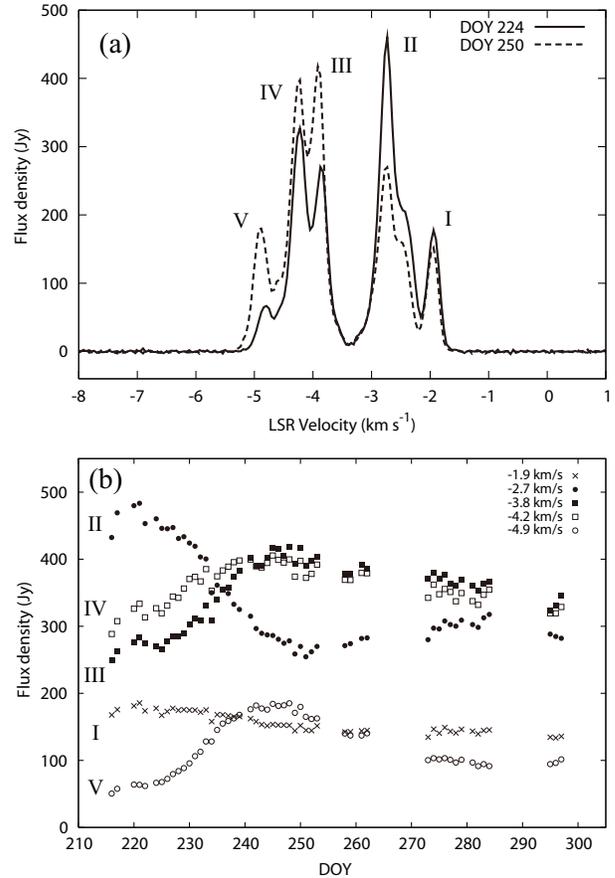}
\end{center}
\caption{The 6.7~GHz methanol maser in Cep~A.
The labels I, II, III, IV, and V
correspond to each spectral feature in each panel.
(a) The spectra obtained in the monitoring observations.
The solid and dashed line shows the spectra
obtained in DOY 224 and 250 observation, respectively.
(b) Time variation of each spectral feature.}
\label{fig:fig1}
\end{figure}

We detected a rapid variability in the spectrum of 6.7~GHz
methanol maser of Cep~A.
Five spectral features, at radial velocity of $-$1.9,
$-$2.7, $-$3.8, $-$4.2, and $-$4.9~km~s$^{-1}$,
were identified during the whole observing period
as shown in figure~\ref{fig:fig1}.
The maser features were labeled as I, II, III, IV, and V,
respectively.
The feature II showed a shoulder at $-$2.5~km~s$^{-1}$,
corresponding a second component.
Since we could not clearly distinguish the two components
of feature II, the second component is not discussed in this paper.

The maser features are divided into two groups;
two red-shifted features (I, II),
and three blue-shifted features (III, IV, V).
These two groups are clearly separated in the spectrum.
They also exhibited different trends of variation
as shown in figure~\ref{fig:fig1}b.
For the first 30~days of the monitoring,
the red-shifted group decreased to 50{\%}
of its initial value (feature II),
while the blue-shifted group increased up to 300{\%} (feature V).
Correlation coefficients for all combinations of spectral features
for the data from DOY 216 to 250 is shown in table~\ref{tab:tab1}.
The absolute values of correlation coefficients are larger than 0.80,
with only exception of 0.63 for I vs IV.
The variation shows a strong correlation within each group,
and an anti-correlation between two groups.

A trend of the variation suddenly changed at around DOY 250.
The red-shifted group started to increase
and the blue-shifted group started to decrease.
The rate of variation was smaller than that of previous period.
This change in the variation trend was clearly synchronized.

\begin{figure*}[htbp]
\begin{center}
\FigureFile(160mm,160mm){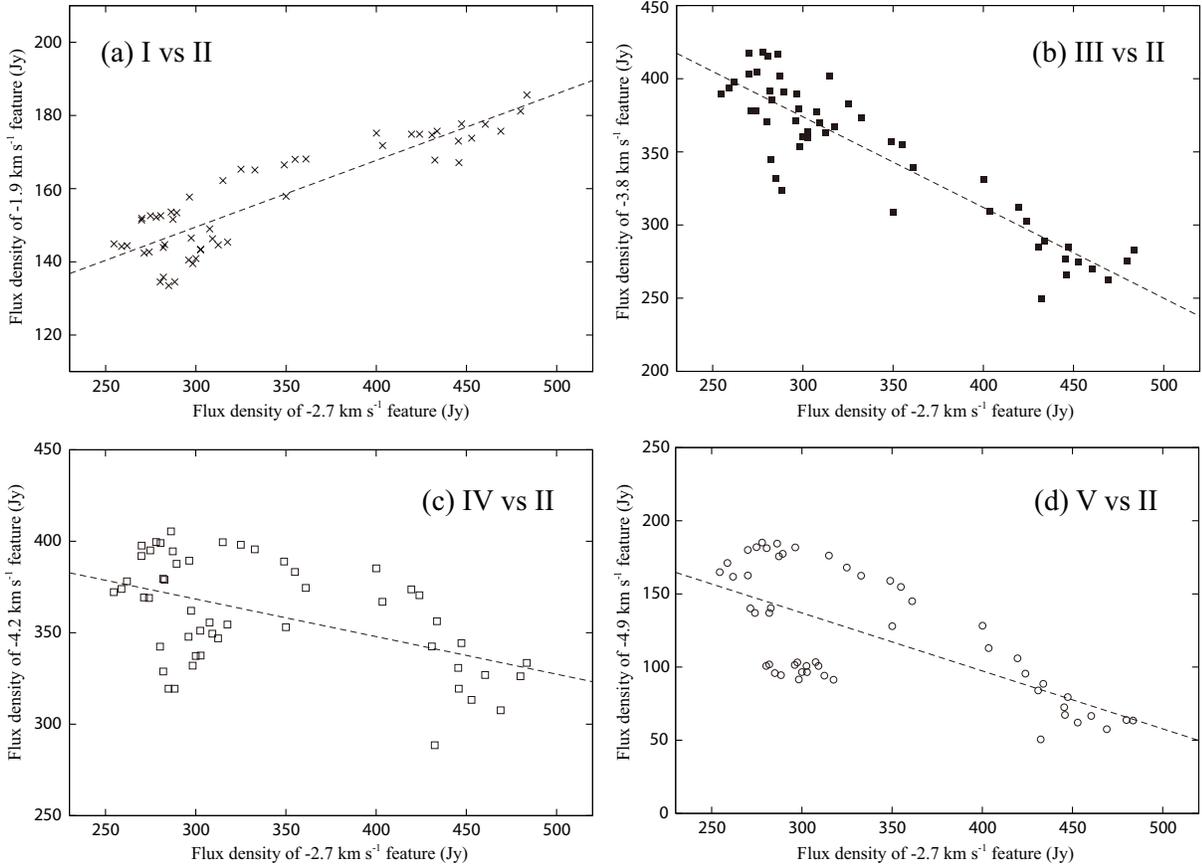}
\end{center}
\caption{The correlation plots with respect to
the spectral feature II.
The symbols in each panel correspond to the spectral features
in figure~\ref{fig:fig1}b.
The dashed lines in each panel
indicate the best fits to the data.}
\label{fig:fig2}
\end{figure*}

Correlation coefficients for the whole period
are shown in table~\ref{tab:tab2}.
The coefficients show essentially the same trends
with those for the first period.
The feature II shows
a strong and positive correlation (0.89) with the feature I,
while anti-correlation was seen with the features III, IV, and V.
The correlation coefficients are $-$0.91, $-$0.50,
and $-$0.68, respectively.
Figure~\ref{fig:fig2} shows the correlation plots for
combinations with respect to the feature II.
The positive (I) and anti-correlation (III, IV, V) is obvious.

Flux variation of feature IV and III are quite similar each other,
but the variation of feature IV is slightly advanced
to that of feature III.
Cross-correlation of the time series of the features
showed small time-delays.
The largest delay is 6~days for II to IV,
and the smallest is 1~day for II to III and V to IV.
We ignored the feature I because of its small variation.

\begin{table}[htbp]
\begin{center}
\caption{Correlation coefficients
from DOY 216 to 250}\label{tab:tab1}
\begin{tabular}{crrrr}
\hline\hline
feature   & \multicolumn{4}{c}{feature} \\ \cline{2-5}
   & \multicolumn{1}{c}{I} & \multicolumn{1}{c}{II}
& \multicolumn{1}{c}{III} & \multicolumn{1}{c}{IV} \\ \hline
II        &   0.94  &         &        &           \\
III       & $-$0.83 & $-$0.96 &        &           \\
IV        & $-$0.63 & $-$0.84 & 0.92   &           \\
V         & $-$0.83 & $-$0.97 & 0.98   &  0.94     \\ \hline
\end{tabular}
\end{center}
\end{table}
\begin{table}[htbp]
\begin{center}
\caption{Correlation coefficients for all days}\label{tab:tab2}
\begin{tabular}{crrrr}
\hline\hline
feature   & \multicolumn{4}{c}{feature} \\ \cline{2-5}
          & \multicolumn{1}{c}{I} & \multicolumn{1}{c}{II}
& \multicolumn{1}{c}{III} & \multicolumn{1}{c}{IV} \\ \hline
II        &   0.89  &         &        &           \\
III       & $-$0.65 & $-$0.91 &        &           \\
IV        & $-$0.08 & $-$0.50 & 0.77   &           \\
V         & $-$0.28 & $-$0.68 & 0.84   &  0.92     \\ \hline
\end{tabular}
\end{center}
\end{table}

\subsection{Spatial Distribution}\label{section:distri}
\begin{figure*}[htbp]
\begin{center}
\FigureFile(140mm,140mm){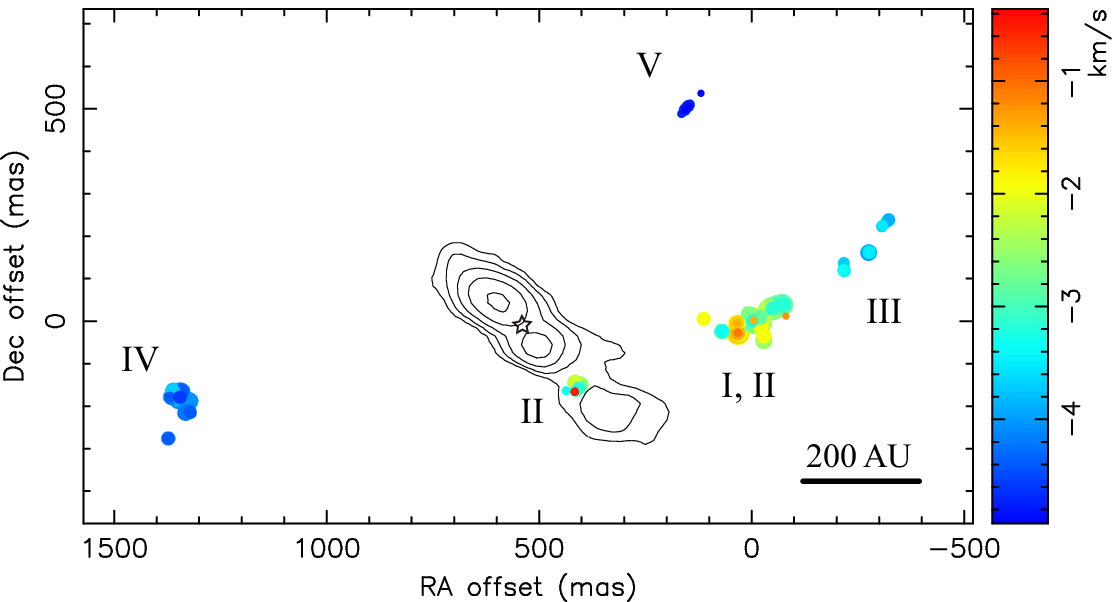}
\end{center}
\caption{{\footnotesize
A spatial distribution of the 6.7~GHz methanol
maser spots (filled circle) of Cep~A.
The spot size and color indicates its
peak intensity in logarithmic scale
and its radial velocity (see color index at the right), respectively.
The contours indicate the VLA 22~GHz continuum
observed by \citet{1998ApJ...509..262T} and re-reduced
by \citet{2003ApJ...586..306G}.
A star indicates the peak of 43~GHz continuum emission
with the positional uncertainty of about 10~mas.
It is thought as the location of an exciting source
\citep{2006ApJ...638..878C}.
The origin of this map corresponds to the absolute coordinate
of the 6.7~GHz methanol maser
($\alpha$(J2000.0)~$=$~\timeform{22h56m17s.90421},
$\delta$(J2000.0)~$=$~$+$\timeform{62D01'49".5769})
obtained by our observation.}}\label{fig:fig3}
\end{figure*}

With the VLBI observation,
117 spots of the methanol maser emission were detected.
Peak intensities of the spots ranged from
$\sim$650~mJy~beam$^{-1}$ to 122~Jy~beam$^{-1}$,
while the rms of image noise (1$\sigma$) in a line-free channel
was 160~mJy~beam$^{-1}$.
The correlated flux accounted for over 90{\%}
of the single-dish flux.
The spatial distribution of the maser spots
(figure~\ref{fig:fig3}) showed an arclike structure.
The size from edge to edge of the arclike structure
is $\sim$1900~mas or $\sim$1400~AU at a distance of 725~pc.

The absolute coordinate of the spot at $-$2.64~km~s$^{-1}$
obtained in our observation is
$\alpha$(J2000.0)~$=$~\timeform{22h56m17s.90421},
$\delta$(J2000.0)~$=$~$+$\timeform{62D01'49".5769},
with errors less than 1~mas.
This is the origin of the image.
The peak of 43~GHz continuum emission (star symbol),
which may be an exciting source \citep{2006ApJ...638..878C},
located near the center of the arclike structure
of the 6.7~GHz methanol maser spots.
The elongation of the arclike structure is nearly
perpendicular to the radio jet.
The overall distribution of the maser spots coincide with
the CH$_3$CN and NH$_3$ disks
(\cite{2005Natur.437..109P}; \cite{2007ApJ...666L..37T})
and the velocity range of the spots is similar to that of these disks,
although a simple velocity gradient was not detected
with the maser spots.
The water maser disk reported by \citet{1996ApJ...457L.107T}
locates almost the same position with the methanol arclike structure,
although the size of the water maser disk is about 2 times smaller.
The ground-state hydroxyl masers around the HW2 object
(\cite{1992MNRAS.254..501M}; \cite{2005MNRAS.361..623B}),
whose internal proper motions were mainly directed away
from the central source,
are distributed surrounding the methanol maser distribution.
The radial velocities of both masers
(water: $-$27.3 to $+$8.9~km~s$^{-1}$;
hydroxyl: $-$25.2 to $-$0.6~km~s$^{-1}$)
cover the range of the methanol maser
($-$4.93 to $-$0.36~km~s$^{-1}$).

The cluster of the spots locating near the origin of the VLBI map
corresponds to the spectral features I and II,
and a cluster locating at \timeform{0".45} east,
\timeform{0".10} south from the origin corresponds to the feature II.
The correspondences of the other clusters and the spectral features
are as follows;
the western cluster (\timeform{0".30} west,
\timeform{0".20} north) is III,
the eastern cluster (\timeform{1".35} east,
\timeform{0".20} south) is IV,
and the northern cluster (\timeform{0".10} east,
\timeform{0".50} north) is V, respectively.
The clusters II (\timeform{0".45} east,
\timeform{0".10} south) and V were detected for the first time
in VLBI observations.
Some weak spots with radial velocities from $-$0.53
to $-$0.36~km~s$^{-1}$, which are a part of the cluster II,
were detected only in the VLBI observation, and not detected
in the spectral observation with Yamaguchi 32-m telescope.
The red-shifted (I, II) and the blue-shifted features (III, IV, V)
are isolated in the spatial distribution by more than 100~AU,
and the red-shifted features are surrounded
by the blue-shifted features.

%%% ******************** %%%
%%% ***  Discussion   ** %%%
%%% ******************** %%%

\section{Discussions}\label{section:discussion}
The time variation of the methanol maser features of Cep~A
was synchronized, and (anti-)correlated.
The synchronized variation were occured at
spatially isolated maser features.
The separation of clusters is larger than 100~AU
in the overall spatial distribution of $\sim$1400~AU,
and it is not likely that there are some mechanical
interactions between features that causes synchronized variation.
The synchronized variation may be caused
by shock wave from a central star,
that is arrived at the clusters simultaneously.
If we assume an outflow velocity of 4.5~km~s$^{-1}$
which corresponds to a radial velocity range
of the Cep~A methanol maser,
it takes more than 1000~years to propagate
through the spatial scale of 1000~AU.
The synchronized arrival to the separated clusters
with a time delay less than several days requires a very fine tuning
of arrival time (6~days/1000~years~$\sim$0.002{\%}),
which is highly unlikely.
The synchronization, the spatial isolation
and the spectral separation of the maser features
suggest that it is difficult to excite the maser
by collision in shock wave from a common exciting source.

These properties favor the widely accepted excitation model
that the 6.7~GHz methanol masers is excited by infrared radiation
from nearby warm dust (e.g., \cite{1997MNRAS.288L..39S}).
Infrared radiation from one variable exciting source
is easy to explain the synchronized variation
with the spatial isolation and the spectral separation.
The exciting source might correspond to that one showed
by \citet{2006ApJ...638..878C}.
The bolometric luminosity of the Cep~A region
is about 2.5~$\times$~10$^{4}$~{\LO} \citep{1981ApJ...244..115E}.
Assuming that half of this luminosity is attributed to the HW2 object
(\cite{1994ApJ...430L..65R}; \cite{1995MNRAS.272..469H}),
we derive a dust temperature of 110~K
at 700~AU from the exciting source.
The distance 700~AU is that from the supposed exciting source
to the farthest maser spot.
This temperature 110~K is consistent with the suitable temperature
($\sim$100-200~K) of regions producing 6.7~GHz methanol maser
(e.g., \cite{2005MNRAS.360..533C}).
This time variation model of one exciting source
may be applicable to the periodic variations of the 6.7~GHz
methanol maser for G9.62$+$0.20E.

The time variation of the Cep~A methanol maser showed anti-correlation
between the red-shifted and the blue-shifted features.
A light-crossing time of $\sim$1400~AU is 8~light-day.
It is consistent with the time-delays (1-6~days)
in the cross-correlation of the flux variation of spectral features.
However, the light-crossing time is too short to cause
the anti-correlation with time-scale more than 20~days.
The anti-correlation could be understood
in terms of excitation environment.
The 6.7~GHz methanol maser requires
a suitable temperature range ($\sim$100-200~K).
Dust temperature is high at regions close to the exciting source,
and is low at regions far from the source.
If the exciting source increases its luminosity,
the temperature near the source would be too high
to produce the maser emission,
while the temperature far from the source
would be suitable to produce the maser.
Although the 3-dimensional distribution
of maser features is uncertain,
the 2-dimensional distribution of the maser features of Cep~A
indicates that the red-shifted features are close to the supposed
exciting source, and the blue-shifted features are far from it.
With this spatial distribution, the anti-correlation of variation
could be explained by the dust temperature change caused
by variability of the exciting source,
i.e., during the period between DOY 216 and 250, the dust temperature
at feature II, which is closer to the exciting source,
increases beyond 200~K and out of the appropriate range
for maser excitation, while the temperature at feature III, IV, and V
increased from 110~K,
which is nearly the minimum value of maser excitation,
to produce stronger maser emission.

Additional observations are required to confirm this model,
in particular to test if short time variation of luminosity occurs
in the exciting source.
The distribution of maser features is observed as
a 2-dimensional projection.
The velocity field would be a clue to understand the 3-dimensional
structure of this region.
We are conducting the VLBI monitoring observations
to detect the internal proper motion of methanol maser spots,
and the results will be reported in future.

%%% ******************** %%%
%%% ***  Conclusion   ** %%%
%%% ******************** %%%

\section{Conclusion}\label{section:conclusion}
We have detected a rapid variability of 6.7~GHz
methanol maser of Cep~A.
Five spectral features are grouped into the red-shifted ($-$1.9 and
$-$2.7~km~s$^{-1}$) and the blue-shifted ($-$3.8, $-$4.2,
and $-$4.9~km~s$^{-1}$).
These two groups are clearly separated in the spectrum.
They also exhibited different trends of variation.
The time variation of this maser features
showed two remarkable properties;
synchronization and anti-correlation
between the red-shifted and the blue-shifted features.
The spatial distribution of the maser spots obtained by
the VLBI observation showed that maser spots of each feature
are largely separated.
The synchronization, the spectral
and the spatial isolation suggest that
the collisional excitation by shock wave from a common
exciting source is unlikely.
We discussed that the time variation of the Cep~A methanol maser
could be produced by one variable exciting source and
this maser could be excited by radiation from the exciting source.
It is possibly thought that the anti-correlation is caused
by a variation of the dust temperature of the maser regions
as a result of the distance to the exciting source,
and the time variation of the exciting source's luminosity.

\bigskip

%%% Acknowledgments

The authors wish to thank the JVN team
for observing assistance and support.
The authors also would like to thank the anonymous referee
for many useful suggestions and comments, which improved this paper.
The JVN project is led by the National Astronomical Observatory
of Japan~(NAOJ) that is a branch of the National Institutes of
Natural Sciences~(NINS), Hokkaido University, Tsukuba University,
Gifu University, Yamaguchi University, and Kagoshima University,
in cooperation with Geographical Survey Institute~(GSI),
the Japan Aerospace Exploration Agency~(JAXA), and the National
Institute of Information and Communications Technology~(NICT).

% \clearpage

\end{document}